This manuscript has been authored by UT-Battelle, LLC, under Contract No. DE-AC0500OR22725 with the U.S. Department of Energy. The United States Government retains and the publisher, by accepting the article for publication, acknowledges that the United States Government retains a non-exclusive, paid-up, irrevocable, world-wide license to publish or reproduce the published form of this manuscript, or allow others to do so, for the United States Government purposes. The Department of Energy will provide public access to these results of federally sponsored research in accordance with the DOE Public Access Plan (http://energy.gov/downloads/doe-public-access-plan).

# Deep Learning Analysis of Defect and Phase Evolution During Electron Beam Induced Transformations in WS$_2$

Artem Maksov[1,2,3], Ondrej Dyck[1,2], Kai Wang[1], Kai Xiao[1], David B. Geohegan[1], Bobby G. Sumpter[1,4], Rama K. Vasudevan[1,2], Stephen Jesse[1,2], Sergei V. Kalinin[1,2,*], Maxim Ziatdinov[1,2,*]

[1]*Center for Nanophase Materials Sciences, Oak Ridge National Laboratory, Oak Ridge TN 37831*
[2]*Institute for Functional Imaging of Materials, Oak Ridge National Laboratory, Oak Ridge TN 37831*
[3]*Bredesen Center for Interdisciplinary Research, University of Tennessee, Knoxville, Tennessee 37996*
[4]*Computational Sciences & Engineering Division, Oak Ridge National Laboratory, Oak Ridge TN 37831*

**Abstract**

Understanding elementary mechanisms behind solid-state phase transformations and reactions is the key to optimizing desired functional properties of many technologically relevant materials. Recent advances in scanning transmission electron microscopy (STEM) allow the real-time visualization of solid-state transformations in materials, including those induced by an electron beam and temperature, with atomic resolution. However, despite the ever-expanding capabilities for high-resolution data acquisition, the inferred information about kinetics and thermodynamics of the process and single defect dynamics and interactions is minima, due to the inherent limitations of manual *ex-situ* analysis of the collected volumes of data. To circumvent this problem, we developed a deep learning framework for dynamic STEM imaging that is trained to find the structures (defects) that break a crystal lattice periodicity and apply it for mapping solid state reactions and transformations in layered WS$_2$ doped with Mo. This framework allows extracting thousands of lattice defects from raw STEM data (single images and movies) in a matter of seconds, which are then classified into different categories using unsupervised clustering methods. We further expanded our framework to extract parameters of diffusion for the sulfur vacancies and analyzed transition probabilities associated with switching between different configurations of defect complexes consisting of Mo dopant and sulfur vacancy, providing insight into point defect dynamics and reactions. This approach is universal and its application to beam induced reactions allows mapping chemical transformation pathways in solids at the atomic level.

\_\_\_\_\_\_\_\_\_\_\_\_\_\_\_\_\_\_\_\_\_\_

\* sergei2@ornl.gov

\* ziatdinovmax@gmail.com

**Introduction.**

Chemical reactions and phase transformations underpin phenomena ranging from cosmological processes, to the emergence of life on Earth, to modern technological progress, and are therefore of tremendous interest for both basic and applied sciences. Simple gas phase reactions of small molecules can be readily studied using well-established spectroscopy methods (infrared,[1,2] mass,[3] NMR[4]), utilizing spatial homogeneity of reaction volumes when the same process occurs multiple times. In conjunction with first-principles-based modelling,[5,6] a reliable picture of molecular reactivity is being built. For studies of more complex organic and biochemical reactions, time-resolved cryogenic microscopy[7] and femtosecond x-ray pump-probes[8] provide a reliable investigative framework, again relying on the statistical similarity between multiple orientations of the same molecule.

The situation is far more complicated for solid state reactions involving continuous solids. Traditionally, solid-state phase transformations and reactions were explored by bulk measurements and x-ray techniques. However, such techniques may not be able to provide sufficient spatial resolution for understanding elementary mechanisms behind the observed transformations. This problem can be partially solved by direct *ex-situ* visualization of reaction zones,[9,10] providing information on the geometry and, in certain cases, atomic configurations at the reaction fronts. Similarly, utilization of colloid models[11] allows for the development of model systems, albeit the nature of local interactions is significantly different from atomic systems.

In recent years, the advances in scanning transmission electron microscopy ((S)TEM) have enabled the direct visualization of dynamic phenomena at the atomic level.[12-20] The physical/chemical phenomena studied by in-situ STEM are wide ranging and now include e-beam induced defect evolution,[21-30] dislocation migration,[31-33] observation of e-beam induced production of single layer Fe and ZnO membranes in graphene nanopores,[34,35] e-beam induced chemical etching and growth from nanoparticle and single atom catalysts,[36-41] sub-10 nm scale lithography,[42] graphene healing,[43] conductive nanowire formation,[44] crystallization and amorphization at 2D interfaces,[45,46] formation of fullerenes,[47] and graphene edge dynamics.[48,49] This list can hardly be considered comprehensive but it serves to illustrate well the vast array of dynamic changes that are being observed and rapidly explored via in situ (S)TEM techniques. A tantalizing development which was published just last year (2017) is the introduction of a single dopant atom into a graphene lattice, the controlled movement of the atom through the lattice, and the assembly of a few primitive structures atom-by-atom.[50-53] Such efforts harken back to the work of Don Eigler who first demonstrated controlled atomic motion via scanned probe techniques.[54] However, given the colorful array of other atomic, chemical process observed in (S)TEM and the continuously growing portfolio of commercially available in-situ equipment (heating, electrical biasing, gas and liquid cells etc.), it seems likely many more processes can be brought under our direct control, turning the (S)TEM into an atomic scale fabrication platform.[55]

Successes in e-beam atom-by-atom fabrication and atom-by-atom mapping of solid state reactions will not only require explorational research and instrumental improvements. The key piece of the puzzle will involve successfully grappling with the enormous amount of data which can be generated by these machines to infer material specific information describing kinetics and thermodynamics of point and extended defects, reaction paths, and mechanisms for extended defect and second phase nucleation and growth. The "by hand" analysis of years past is no longer a tractable solution considering the dimensionality and number of datasets which are now routinely obtained. This necessitates the creation of methods which allow for automated analysis of dynamic transformations to extract relevant materials descriptors and reconstruct reaction pathways from various sources of detector readouts, such as the variety imaging and spectroscopic modes. In this article we attempt to forge an inroad in one aspect of this challenge, namely automated image analysis for the detection and tracking of defects in STEM video of 2D materials and further proceed to extract microscopic point defect reaction mechanisms from these observations.

Here, we analyze the phase evolution of Mo-doped $WS_2$ during electron beam irradiation. In this process, the electron irradiation results in removal of the sulfur atoms, rendering the system oversaturated with respect to low-valence tungsten-sulfur moieties. We develop a deep learning network for rapid analysis of this dynamic data, analyze transformation pathways, create a library of defects, and explore minute distortions in local atomic environment around the defects of interest, ultimately building a complete framework for exploring point defect dynamics and reactions.

**Results and Discussion.**

Figure 1 shows several selected frames from the STEM "movie" of lattice transformations in the Mo-doped $WS_2$ monolayer under 100 kV electron beam irradiation. The full "movie" is available in the Supplementary Material. It can be clearly seen that the system evolves with time, evolving numerous point defects. On accumulation of non-stoichiometry, the latter start to segregate, forming extended defect, nucleating secondary phases, and resulting in the segmentation and rearrangement of the 2D layer. The key task is get out the information of interest about the defects. Unfortunately, most of the methods for localizing and identifying/classifying defects available to date are slow, inefficient and require frequent manual inputs.

To overcome the limitations of the available approaches, we developed physics-based machine learning method for localizing and identifying defects. We exploit the fact that each defect is associated with violation of ideal periodicity of the lattice. Therefore, we train a convolutional neural network (cNN) using a *single* image at the early stage of the beam-induced transformation, when macroscopic periodicity is still maintained, and each defect can be readily discovered providing the "ground truth" for network training. Thus, a trained network relies only on the local characteristics of the image, and hence can identify defects on the later stages of system evolution when the long range periodicity of the lattice is broken due to a second phase evolution and displacement and rotation of unreacted $WS_2$ fragments. Furthermore, we find that the network can discover via "extrapolation" other defects which may not necessarily be a part of the initial training set. Such "extrapolation" is possible due to generalization abilities of deep learning models. Indeed, we have recently demonstrated[56] that a deep cNN trained on the simulated images of an idealistic lattice vacancy structure can in principle generalize well enough to detect larger and more complex lattice vacancy structures in the system (*e.g.*, double and triple vacancies, as well as reconstructed vacancies). The extracted defect structures can be identified/classified using unsupervised clustering and unmixing techniques. Finally, the selected defects can be studied further using local crystallography techniques,[57] such as a combination of atom finder and principal component analysis for analyzing minute atomic distortions in their vicinity in the "movies", as well as with a Markov analysis for identifying transition probabilities between different defect configurations.

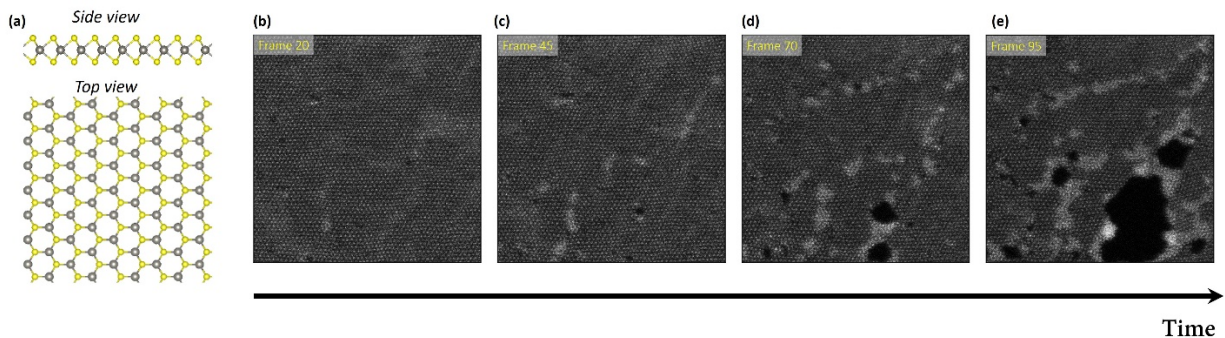

**FIGURE 1. Defects evolution under e-beam irradiation in Mo-doped $WS_2$.** (a) Ball-and-stick representations of $WS_2$ structure. (b-e) Four selected frames from the STEM "movie" of Mo-doped $WS_2$ obtained at 100 kV illustrating formation of defects and lattice transformations as a function of time. The full movie can be found in the supplementary material.

As a first step of analysis, we define the topology of a neural network to target specific physics of beam-induced transformations. The network must be able i) to separate atomic scale lattice disorder from the rest of the lattice, ii) to return the precise location of the detected defects, iii) to be able to generalize to previously unseen defect structures. One possible candidate is the class activation maps based deep learning analysis, in which a model trained on image-level labels is capable in principle of discriminating the image regions used to identify the specific class[58] (defect). The disadvantage of such approach is that one must start with manually selecting the isolated single defect structures to create a training set. In addition, we found that while this approach allows certain atomic defect structures to be located with a sufficient accuracy, it does show relatively poor generalization ability. The alternative approach is to use a fully convolutional neural network model,[59] which can be trained to output a pixel-wise classification map, with the same size/resolution as the original input image, that shows a probability of each pixel belonging to certain type of object (defect). This type of model has been recently successfully applied to finding lattice atoms in raw STEM data[56] and we therefore chose it for the current problem.

The next task is to create a training set that will be used to "teach" a model to find lattice defects in STEM "movies", allowing for sufficient flexibility to discover all the defects but at the same time avoiding over-classification for classes that physically cannot be present in the data. We found that it is possible to train a network using only the first frame of such a "movie" or a single image obtained before recording a movie, and then let the trained network analyze the remaining part of the movie. This approach utilizes the fact that in macroscopically (i.e. on the length scale of the image) system the defects can be trivially discovered via the Fourier method,[60] providing the ground truth for network training. However, when trained, the network relies solely on local edge properties for identification and is thus stable towards formation of extended defects, rotations, and fragmentations of the lattice.

To identify the defects, we select a single image (frame) at the beginning of transformations (Fig. 2a). Once the image is selected, we perform global Fast Fourier Transform (FFT) on the selected experimental image and apply a high pass filter in reciprocal space in order to remove non-periodic components of the lattice (Fig. 2b). We then perform inverse FFT to obtain periodic image and subtract the original image from it (and *vice versa*) such that only the deviations from the ideal periodic lattice remain.[60] In this image difference, vacancies show up as bright spots. Next, the image difference is thresholded to find locations of the single defects (Fig. 2c). Note that the thresholded image represents 'ground truth' which will be used to train a convolutional neural network. The training set is created by performing data augmentation of the selected experimental image and the corresponding ground truth image. This augmented dataset can be used to train a neural network to return positions of atomic lattice disorder from raw experimental data (Fig. 2d). Once trained, not only this cNN-based method for finding defects is faster and more efficient than the method based on FFT subtraction, but it also allows, unlike the FFT method, to find position of defects in the images of fragmented atomic lattice where multiple (joint and/or disjoint) lattice domains can be rotated by different angles with respect to each other. Because our model allows finding defects that break lattice periodicity irrespectively of the exact type of the defect, we consider it to be a 'universal' defect finder for a given material.

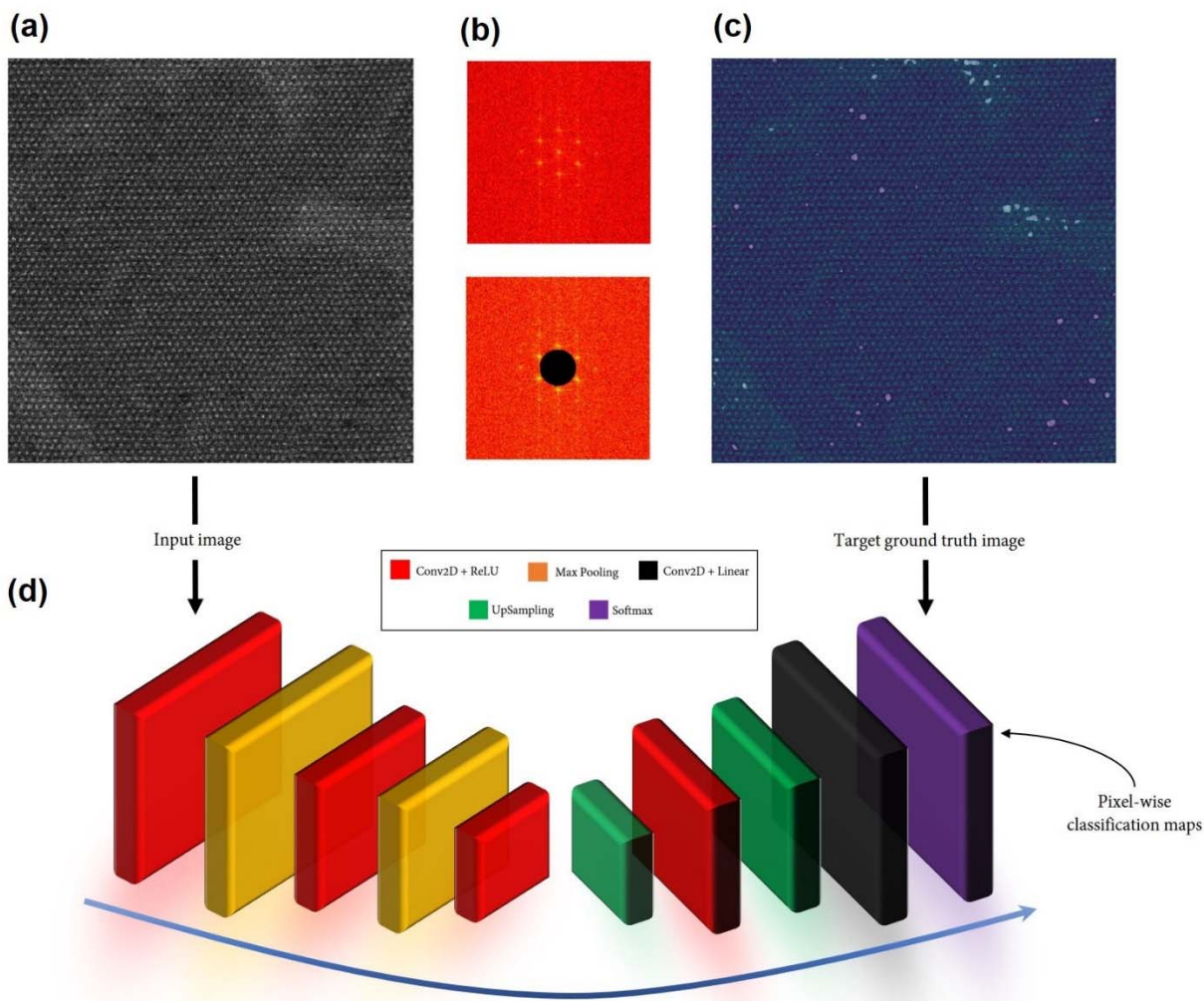

**FIGURE 2. Training deep convolutional neural network to recognize defects that break lattice periodicity.** (a) The first frame from STEM "movie" on Mo-doped $WS_2$. (b) Global FFT and global FFT with high-pass filter applied. (c) Binary masks for image differences between the original data in (a) and inverse of filtered FFT in (b). The image in (a) is a training image and the data in (c) serves as 'ground truth' (pixel-wise labeling). (d) Schematics of convolutional neural network with an encoder-decoder type of structure.

We now use the cNN model trained according to the method described above (accuracy on the test set ~99 %) to locate atomic defects in dynamic STEM data on Mo-doped $WS_2$. Interestingly, although our model was trained using only the 1$^{st}$ frame (out of 100) of the movie, it was able to accurately identify the positions of atomic defects in the remaining 99 frames (see Supplementary Figure 1; full movie of the defects found can be downloaded from Supplementary Material). Once a sufficient number of defects (~$10^4$ in this case) is extracted via the cNN model it becomes possible to categorize them into different classes. To perform such a defect classification in unsupervised fashion, we adapted a Gaussian mixture model (GMM).[61] The GMM is applied to a stack of defect "windows" (images of the identical size, usually 32px*32px cropped around the center of each defect) extracted using the pixel-wise classification maps in the cNN output. Here, we chose the number of GMM components to be five as it appears to be an optimal number of components for understanding the type of defect structures present in the data. Indeed, an increase in the number of components resulted in fine (sub-) structures of the detected defects, while decrease in the number of components produced some physically meaningless structures (see Supplementary Figure 2). We also note that the number of components past purely exploratory stage can be adjusted based on known defect chemistry of material (either from common physics principles, DFT calculations, or combinatorial analysis)

The defect structures associated with the unmixed components of GMM are shown in Fig. 3a-e. The class 1 and class 3 (Fig. 3a,c) were found to correspond to a substitutional atom in W sublattice with a lower Z number, which we interpret as Mo dopant (**Mo$_w$**). Note that Mo atom does not occupy a symmetric central spot in these structures as one would expect for a lone Mo dopant. This suggests that there are additional distortions present in the defects that form classes 1 and 3, likely associated with a disorder in S sublattice. Interestingly, presence of a coupling between distinct defect species has been recently observed in static STEM images from Mo-doped WS$_2$ system and attributed to merging of defects during growth and post-growth procedures.[62] This comparison illustrates that as in other cases, systematic application of statistical and machine learning methods allows both to recover earlier observations and, as we show next, derive new information about underlying physical and chemical processes. The class 4 and class 5 (Fig. 3d, e) are associated with vacancy in W sublattice (**V$_w$**) and in S sublattice (**V$_s$**), respectively. Presence of adatoms / "contaminations" created during the e-beam surface transformations (e.g. chemical species from initial WS$_2$ material deposited back on to the surface in combination with carbon atoms) can explain a defect structure from the class 2 (Fig. 3b).

Figure 3f shows spatio-temporal trajectories ('brush diagram'[63]) for the identified defects. Based on the analysis of the diagram, we identify three characteristic statistical behaviors: weakly moving trajectories, stronger diffusion, and "uncorrelated events" / "flickering". Presence of more than one characteristic behavior of the atomic defects may be potentially connected to complex spatial character of strain fields during the material transformation, which may impact diffusion properties as well as create certain "localization regions", in which the motion of defects is restrained.[64,65] In the following we will focus on the analysis of the continuous and quasi-continuous trajectories only. The most well-defined trajectories are associated with Mo dopants (class 1 and class 3). These Mo defects show different diffusion behaviors depending on their location in the lattice and are characterized by reversible switching between two configurations (class 1 and class 3) along their trajectories. The defects associated with S and W vacancies typically form shorter (compared to Mo defects) trajectories. One possible explanation is that these vacancies are getting filled by the W and S species from the extended clusters of the deposited $WS_2$ material (although we did not find any associated correlations with point defects of class 2).

We now demonstrate that, based on the results produced by a combination of cNN and GMM, it becomes possible to estimate diffusion characteristics of the selected defect species. Particularly, we studied diffusion properties of S vacancies. We first collapse the 3-d spatiotemporal diagram for a chosen class of defect into the 2-d representation. For this purpose, we project the 'windows' of specific class of defect, which allows to separate defects that are continuous in time from the randomly occurring ones (see Supplementary Figure 3). This analysis is complemented by a density-based clustering algorithm,[66] which yields similar results. After extracting defect coordinates for each selected defect "flow" (Fig. 4), we can obtain variance of each distribution and estimate a diffusion coefficient within a framework of a random walk model in two dimensions. This yields values of diffusion coefficient between $3\times10^{-4}$ $nm^2/s$ and $6\times10^{-4}$ $nm^2/s$.

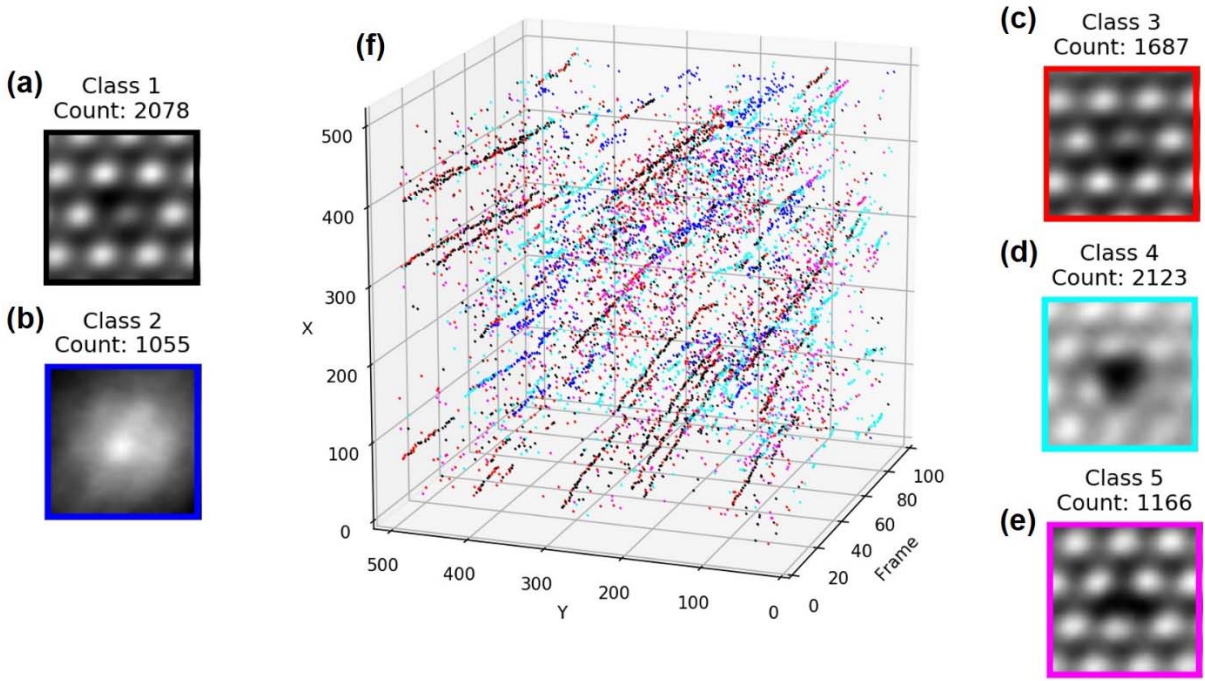

**FIGURE 3. Unsupervised classification of defects located by the deep convolutional neural network.** (a-e) Results of GMM unmixing into 5 components (classes). (f) Spatio-temporal trajectories of the detected defects. Color scheme is the same as in (a-e).

We further proceed to the analysis of another type of defect, namely, the defect associated with Mo dopant (class 1 and 3). Here it is worth noting that while the GMM-based decomposition into 5 components provides us with a good understanding of the major types of defects present in the system, it doesn't allow studying the fine details (variations) of the detected structures. Performing such an analysis is especially important for classes 1 and 3 that show peculiar switching behavior in Fig. 3f. We therefore investigated the "internal" structures of classes 1 and 3 using the so-called local crystallography analysis.[57] Specifically, we studied statistically significant deformation of the nearest neighborhood for each defect structure using principal component analysis (PCA). We first employ a deep-learning-based 'atom finder'[67] that allows extracting positions of atoms from thousands of noisy images of defects in a matter of seconds (note that S atoms cannot be reliably identified at the current experimental resolution and hence we omit them). The first 2 PCA components associated with displacements from the averaged structure of the central Mo atom and six W neighbor atoms for each defect class are plotted in Fig. 5a,b. Since Mo dopant does not considerably distort the $WS_2$ lattice,[62] the structural variations in PCA analysis must be associated with a disorder in S sublattice. In general, one must exercise caution in assigning a specific physical meaning to the PCA components. However, the results shown in Fig. 5 strongly suggest a presence of strong variations in a relative position of central Mo atom with respect to neighbor W atoms, thus it is possible that these variations originate from the presence of S vacancies next to Mo dopant.

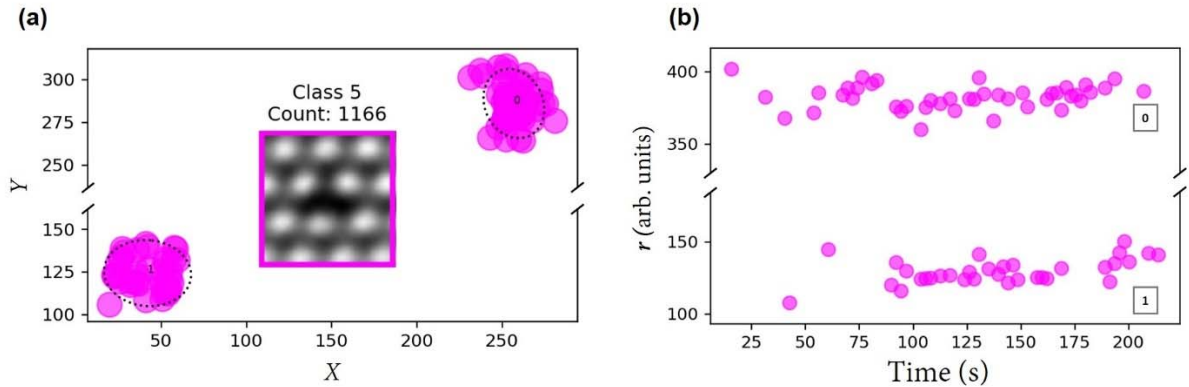

**FIGURE 4. Analysis of diffusion behavior for selected defect structure.** (a) 2-d projections (*X-Y* coordinates) of the 3-d defect "flow" of the S vacancies (inset) with 95% prediction ellipses overlaid. (b) 1-d *r(t)* representation of the same data.

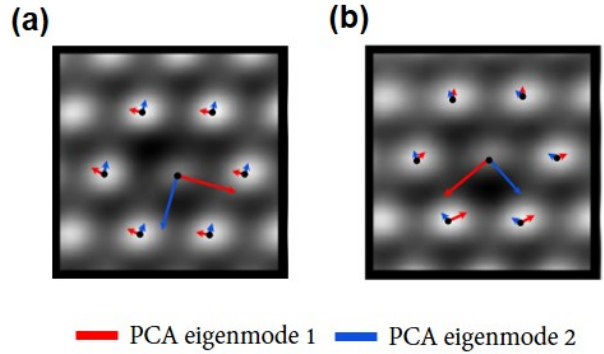

**FIGURE 5. Local crystallography analysis for the selected defect classes.** (a, b) The principal component analysis-derived first two eigenmodes of atomic displacements for defects associated with a Mo dopant (class 1 and class 3 in Fig. 3a and 3c, respectively) presented as vectors of deformation from atom positions in the averaged defect structure. The length of the arrows is magnified by a factor of 5.

Based on the PCA analysis of the atomic displacements in Fig. 5a,b and general lattice symmetry considerations we use GMM to split the defect structures from classes 1 and 3 into four subclasses (Fig. 6b) associated with undistorted $Mo_w$ defect (no coupling to S vacancy) and three ($Mo_w + V_s$) complexes (it is worth noting that the similar result can be achieved by splitting the entire stack of all the defect images into >12 classes, see Supplementary Figure 2). Our next goal is to analyze the switching behavior between different states. Using the same approach as described for the analysis of diffusion parameters we first identified continuous in time defect trajectories for all the 4 subclasses from Fig. 6b, isolated them, and then converted into $r(t)$ 1-d representation (Fig. 6a). In this case, each "flow" represents a sequence (in time) of defect structures undergoing a switching between four different states. This observation suggests that the switching between different states can be analyzed as a Markov process, defining corresponding reaction constants on a single defect level.

The corresponding Markov transition matrix is depicted in Fig. 6d (see also the schematics of transitions in Fig. 6c). This analysis suggests the $Mo_w$ defect may couple to S vacancy in the dynamic STEM experiment. To explain transitions between $Mo_w$ and ($Mo_w + V_s$) we argue that, due to a lower diffusion barrier of a S vacancy, as well as higher probability of S sublattice atoms being knocked-out during the e-beam irradiation, it is likely that the S vacancy created in the vicinity of Mo dopant can get "captured" by it. Interestingly, we also found transitions between different ($Mo_w + V_s$) structures. While the detailed explanation of such a behavior would require rigorous first-principles calculations and additional experiments, one can argue that the ($Mo_w + V_s + V_s$) structures are not stable and/or have a very short "lifetime" compared to the experimental time resolution (to the best of our knowledge such structures have not been observed even when (**Mo$_w$ + V$_s$)** defects are abundant) and therefore attachment of the second S vacancy leads to "pushing" one of the two S vacancies out of the structure. The noticeably different values of transition probabilities for (**Mo$_w$ + V$_s$) - I** structure can be explained by a different rate of "supply" of S vacancies from different lattice directions, for example due to non-trivial distribution of strain fields during e-beam induced transformation and their effect on diffusion characteristics in different lattice directions.

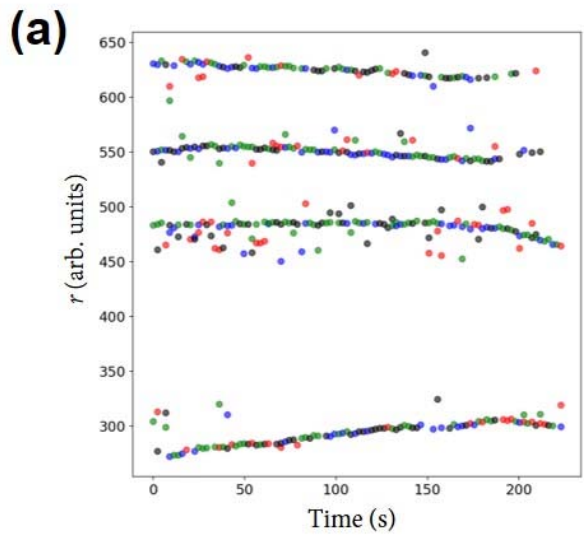
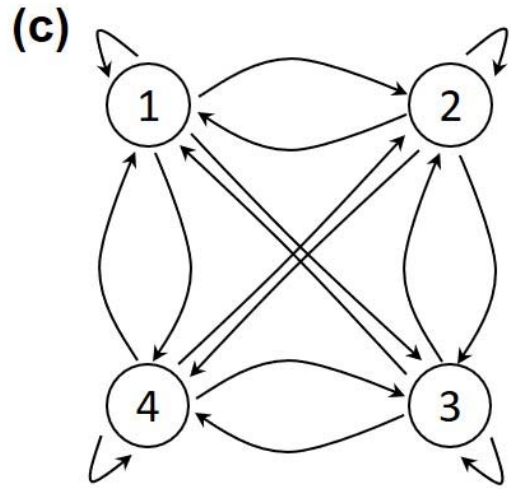
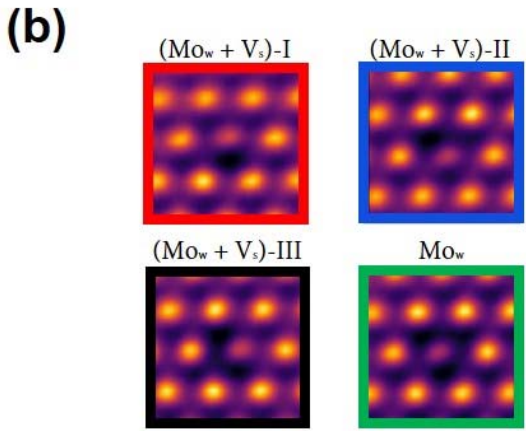
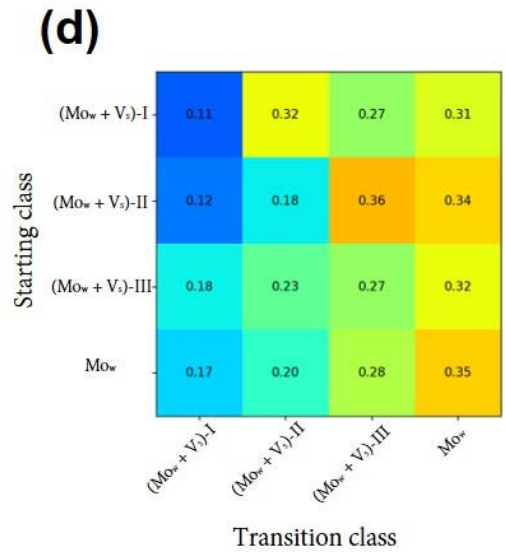

**FIGURE 6. Analysis of transition probabilities between different defect states.** (a) 1-d representation of defect "flows" for the 4 subclasses of defect associated with Mo dopant in (b). Color scheme is the same as in (a) and (b). (c) Illustration of Markov transition processes between 4 states. (d) Markov transition matrix for the 4 subclasses (lone Mo dopant and 3 complexes of Mo dopant with S vacancy) based on the analysis of trajectories in (a).

**Conclusions and Outlook.**

In summary, we have presented a deep-learning-based approach for analysis of dynamic transformation of the lattice structure in STEM "movies" from Mo-doped $WS_2$. We started by teaching a deep neural network how the defects that break lattice periodicity appear in STEM data using a single experimental image (frame 0) and then used the generalization abilities of the network to find various types of atomic defects in the rest of the experimental data. We then performed unsupervised classification of the detected defect structures using Gaussian mixture model and showed that the classification results can be linked to specific physical structures. We were then able to i) identify dominant point defects and their characteristic statistical behaviors in the spatiotemporal diagrams, ii) analyze diffusion for the selected defect species (S vacancy), and iii) study transformation pathways for Mo-S complexes, including detailed transition probabilities. In this manner, point-defect dynamics and solid state reactions in material are studied on atomic level, and corresponding reaction constants are determined for just one point defect.

As far as the future studies are concerned, we believe that one particularly promising direction is incorporating specific physics-based constraints into the machine learning based analysis of STEM videos. Indeed, the current approaches treat observed lattice defects as collections of pixels, without "understanding" the physics behind the observations. One possible way of overcoming such physics-agnostic classification is by integrating a Markov model into the initial search and identification/classification scheme. The Markov model can be guided by the theoretical calculations of interaction potentials on the atomic level, enforcing physical constraints to transition probabilities of atoms and molecules, effects of electron beam on the matter, operating both in space (hidden Markov random field) and time (hidden Markov model) domains. For example, one may incorporate transition probabilities between certain types of defects (*e.g.*, reconstructed vs. non-reconstructed defect), as well as a maximum diffusion length of a defect for a given time scale calculated from first principles, and with a Markov model use it to refine the results of the initial classification. This would be an important step towards creating a fully-autonomous, AI microscope that is making decisions based on the knowledge of physics that it was "taught".

**Data availability**

The complete workflow for studying defects in dynamic STEM data, which includes creation and training/testing of DL model, unsupervised defect classification, analysis of diffusion characteristics, local crystallography analysis and Markov transition matrix analysis, is available in a form of Jupyter notebooks at https://github.com/artemmaksov/ORNL-DeepLearningForAtomicScaleDefectTracking.

**Methods**

**Sample preparation**

The Mo doped $WS_2$ monolayers were grown on $SiO_2$/Si substrate at 800ºC by a low-pressure chemical vapor deposition.[43] To prepare STEM samples, poly(methyl methacylate), PMMA (A4), was first spun onto the $SiO_2$/Si substrate with monolayer crystals at 3500 rpm for 60 s. After being cured at 100 °C for 15 min, the PMMA/$W_{1-x}Mo_xS_2$ sample was detached from the substrate with a 30% KOH solution (100 °C and 0.5−1.0 h). The sample was then transferred to DI water to remove the KOH residue. The washed film was scooped onto a QUANTIFOIL TEM grid. The PMMA was then removed with acetone, and the samples were soaked in methanol for 12 h to achieve a clean surface with flakes. To remove the polymer, the TEM grids were then annealed in an Ar flow (90 sccm, 10 Torr) at 350 °C for 3 h.

**STEM experiment**

STEM imaging was performed using a Nion UltraSTEM U100 STEM operated at 100kV. The images were acquired in high angle annular dark field (HAADF) imaging mode and were introduced to the deep convolutional neural network without any post processing.

**Data Analysis**

The deep convolutional neural network was implemented using Keras 2.0 (https://keras.io) Python deep learning library, with the TensorFlow backend. The convolutional neural network had an encoder-decoder type of structure. The encoder part had alternating convolutional layers for feature extraction with filters of the size 3×3 and stride 1 activated by a rectified linear unit (ReLU) function and max-pooling layers of the size 2×2 and stride 2 to account for translational invariance as well as for reducing the size of processed data. The decoder part of the network, whose role was to map the encoded low-resolution feature maps to full input-resolution feature maps, consisted of the same filters (in reverse order) and upsampling layers. The feature maps from the final convolutional layer of the network were fed into a softmax classifier for pixel-wise classification, providing us with information on the probability of each pixel being a defect.

The Adam optimizer[68] was used for training. The Gaussian mixture model was implemented with scikit-learn machine learning library (http://scikit-learn.org).

**Acknowledgement**

This research was sponsored by the Division of Materials Sciences and Engineering, Office of Science, Basic Energy Sciences, US Department of Energy (RVK and SVK). Research was conducted at the Center for Nanophase Materials Sciences, which is a DOE Office of Science User Facility, and was also supported by the Laboratory Directed Research and Development Program of Oak Ridge National Laboratory, managed by UT-Battelle, LLC, for the U.S. Department of Energy (MZ, OD, SJ). The synthesis of Mo-doped $WS_2$ 2D materials (KW, KX, DG) was supported by the U.S. Department of Energy, Office of Science, Basic Energy Sciences, Materials Sciences and Engineering Division. AM acknowledges fellowship support from the UT/ORNL Bredesen Center for Interdisciplinary Research and Graduate Education.

**Contributions**

M.Z. and S.V.K. conceived the idea. O.D. acquired the STEM data. A.M. and M.Z. wrote the code and analyzed the data. M.Z. and S.V.K. wrote the paper. K.W., D.G., and K.X. were involved in fabrication of the $WS_2$ film. R.V.K., S.J. and B.G.S. aided in the interpretation of results. All authors commented on the manuscript.

**Competing interests**

The authors declare no competing interests.